\documentclass[prl,aps,twocolumn]{revtex4}
\usepackage{graphicx}
\usepackage{amsmath}
\begin{document}
\title{Collision quenching in the ultrafast dynamics of plasmonic materials}
\author{Andrea Marini$^{1}$}
\email{andrea.marini@aquila.infn.it}
\author{Alessandro Ciattoni$^{2}$}
\author{Claudio Conti$^{3,4}$}
\affiliation{$^1$Department of Physical and Chemical Sciences, University of L'Aquila, Via Vetoio, 67100 L'Aquila, Italy}
\affiliation{$^2$Consiglio Nazionale delle Ricerche, CNR-SPIN, Via Vetoio 10, 67100 L'Aquila, Italy}
\affiliation{$^3$Institute for Complex Systems (ISC-CNR), Via dei Taurini 19, 00185, Rome, Italy}
\affiliation{$^4$Department of Physics, University Sapienza, Piazzale Aldo Moro 5, 00185, Rome, Italy}

\begin{abstract}
We explore the nonlinear response of plasmonic materials driven by ultrashort pulses of electromagnetic radiation with temporal duration of few femtoseconds and high peak intensity. By developing the Fokker-Planck-Landau theory of electron collisions, we solve analytically the collisional integral and derive a novel set of hydrodynamical equations accounting for plasma dynamics at ultrashort time scales. While in the limit of small light intensities we recover the well established Drude model of plasmas, in the high intensity limit we observe nonlinear quenching of collision-induced damping leading to absorption saturation. Our results provide a general background to understand electron dynamics in plasmonic materials with promising photonic applications in the manipulation of plasma waves with reduced absorption at the femtosecond time scale.
\end{abstract}
\maketitle

\paragraph{Introduction --}

Plasmonic materials and metamaterials (MMs) - artificial media composed by plasmonic circuits arranged in periodic patterns - have attracted over the last decade a great deal of interest owing to their capacity to confine light down to the nanoscale. Such a peculiar property ensues from the coupling between photons and plasma waves excited within plasmonic media such as metals, doped semiconductors, or graphene, enabling the excitation of surface plasmon (SP) modes \cite{Ritchie1968,Zayats2005}. The tight confinement provided by SP modes in plasmonic media and MMs has opened several possibilities to enhance light-matter interaction and nonlinear optical processes \cite{Kauranen2012}, enabling control and manipulation of light at the nanoscale \cite{Barnes2003} along with the development of nano-sized optical interconnects \cite{Ozbay2006}, super-resolution techniques \cite{Pendry2000,Willets2017}, surface-enhanced spectroscopy \cite{Moskovits1985,Demirel2018}, and optical cloaking \cite{Cai2007,Fleury2015}. 

Most of the above mentioned innovative applications have been developed thanks to the continuous advances in nanofabrication techniques, which have exploited mainly noble metals as plasmonic materials. However, except for sensing and surface-enhanced spectroscopy, the amount of loss in current plasmonic devices such as interconnects, switches, modulators, and detectors is far too high for practical applications. More recently, graphene \cite{Bonaccorso2010,Javier2014}, oxides and nitrides \cite{Naik2011,Kim2013,Naik2013,Guler2015}, and polar dielectrics \cite{Caldwell2015} have risen as promising materials for infrared plasmonics with smaller absorption owing to the lower electron density. Further strategies for loss mitigation involve quenching of interband absorption via self-induced-transparency plasmon solitons \cite{Marini2013}, reduction of surface roughness \cite{Wu2014}, and embedding gaining media in plasmonic setups \cite{Bergman2003,Noginov2008,Noginov2009,Marini2009,Stockman2010,Bolger2010,DeLeon2010}. The ultrafast response of plasmonic media to sub-picosecond temporal pulses of electromagnetic (EM) radiation generally involves heating of electrons and the lattice with a complex dynamics involving exchange of energy from hot electrons to the lattice via electron-phonon scattering \cite{CarpenePRB2006,Rotenberg2007,BaidaPRL2011,ConfortiPRB2012,MariniNJP2013,Khurgin2015}. Heating is a primary consequence of absorption and is generally undesirable since it leads to material damage. Reducing the time-duration of EM pulses to the femtosecond regime is a promising strategy to quench absorption and heating since it enables the achievement of high peak intensity with reduced transfer of energy to hot electrons. 

In this Letter, we investigate the ultrafast dynamics of plasmonic media upon excitation by temporal pulses of EM radiation with duration of few femtoseconds. Starting from the Boltzmann equation for the electron plasma, and accounting for electron-electron and electron-phonon collisions through the full collisional integral, we derive a Fokker-Planck-Landau equation (FPLE) specialized for plasmonic materials under the assumption of weak coupling. Thus, we solve the FPLE through the method of moments obtaining a novel set of hydrodynamical equations (HDEs) describing electron collisions beyond the standard relaxation approximation. We find a complex analytical expression for the effective electron damping that depends nonlinearly on the electron current and reduces to the Drude-like damping in the limit of small driving intensity, while it quenches for high driving peak intensity. By specializing our model to silver, we observe that the intensity of absorption saturation depends quadratically on the EM angular frequency and is of the order of GW$/$cm$^2$ at near infrared frequencies.



\paragraph{Background --} The optical response of a plasmonic material depends mainly on the intraband dynamics of electrons in the conduction band and is determined by the Schr\"{o}dinger equation resulting from the sum of unperturbed and dipole interaction Hamiltonians. In principle, the dipole interaction Hamiltonian can excite interband transitions between valence and conduction bands, which are generally undesirable since they lead to increased resonant absorption. However, when the angular frequency $\omega$ of the driving EM radiation is such that the photon energy $\hbar\omega$ is much smaller than the energy bandgap $E_{ \rm g}$ interband transitions are detuned and become inefficient. Conversely to interband transitions, quantum intraband dynamics of free electrons with density $n$ and effective mass $m$ is fully equivalent to the Drude classical approach, which describes the plasma response in terms of the well-known susceptibility $\chi_{\rm Drude}(\omega) = - \omega_{\rm P}^2/[\omega(\omega+{\rm i}\gamma)]$, where $\omega_{\rm P} = \sqrt{ne^2/\epsilon_0 m}$ is the plasma frequency, $-e$ is the electron charge, $\gamma$ is an effective damping rate, and $\epsilon_0$ is the dielectric permittivity of vacuum. This result is readily obtained also within the classical kinematic approach, where a plasma is modeled as a dilute gas of $N$ electrons enclosed in a box of volume $V$. The electron temperature $T_{\rm e}$ is assumed sufficiently high and their density sufficiently low such that they are localized wave packets with De Broglie wavelength much smaller than their average separation: $\hbar(N/V)^{1/3}/\sqrt{2mk_{\rm B}T_{\rm e}}<<1$. In such conditions, which are fully met in typical optical experiments with plasmonic media, electrons can be considered as non-interacting distinguishable classical particles with charge $-e$ and rather well-defined position ${\bf r}$ and velocity ${\bf w}$.

\paragraph{Classical kinematic theory --} In this framework, the out of equilibrium (OOE) plasma dynamics is modeled in terms of the time dependent distribution function $f({\bf r},{\bf w},t)$, defined in such a way that $f({\bf r},{\bf w},t){\rm d}{\bf r}{\rm d}{\bf w}$ is the number of electrons that, at time $t$, have positions lying within a volume element ${\rm d}{\bf r}$ about ${\bf r}$ and velocities lying within a velocity-space element ${\rm d}{\bf w}$ about ${\bf w}$, so that $\int f({\bf r},{\bf w},t){\rm d}{\bf r}{\rm d}{\bf w} = N$. The time-dependent evolution of $f({\bf r},{\bf w},t)$ is governed by the Boltzmann equation
\begin{equation}
\left( \partial_t + {\bf w}\cdot\nabla_{\bf r} + \frac{1}{m} {\bf F}_{\bf w}({\bf r},t)\cdot\nabla_{\bf w} \right) f = (\partial_tf)_{\rm coll}, \label{BoltzEq}
\end{equation}
where ${\bf F}_{\bf w}({\bf r},t) = - e {\bf E}({\bf r},t) - e {\bf w} \times {\bf B}({\bf r},t)$, ${\bf E}({\bf r},t)$ and ${\bf B}({\bf r},t)$ are the external electric and magnetic fields, and $(\partial_tf)_{\rm coll} = (\partial_tf)_{\rm coll}^{\rm el-el}+(\partial_tf)_{\rm coll}^{\rm el-ph}$ is the collision rate with other electrons and phonons. Within the Boltzmann Stosszahlansatz of molecular chaos \cite{Balescu}, the collision rate is a complex integral expression depending nonlinearly over $f({\bf r},{\bf w},t)$ \cite{Balescu}. The system under consideration is modeled as a dilute gas of free electrons immersed in a homogeneous background of ions with constant density $n_0 = N/V$ (equivalent to the lattice density in the case of a plasmonic solid), mass $M$, ambient temperature $T_0$, and time-independent distribution function $f_0({\bf r},{\bf w})=n_0(M/2\pi k_{\rm B}T_0)^{3/2}{\rm exp}[-Mw^2/2k_{\rm B}T_0]$, such that the total charge vanishes $e \int {\rm d}{\bf r}{\rm d}{\bf w} (f_0 - f) = 0$. Since ions are much heavier than electrons, we neglect their motion and focus only on electron dynamics. Nevertheless, the ion mass density enters the electron-phonon collision rate, which damps free oscillations of the electron plasma and relaxes the electron temperature towards equilibrium. 

\paragraph{Weak coupling --} In general, the temporal solution of the Boltzmann equation - upon excitation by an external EM field - provides the plasma response of the system. However, such an approach does not enable the analytical evaluation of the collision rate and is computationally demanding. For this reason, standard theoretical approaches to evaluate the effect of electron collisions approximate the collision integral by $(\partial_tf)_{\rm coll} \simeq -(f-f_0)/\tau_{\rm eff}$, where $\tau_{\rm eff}$ is an effective relaxation time, typically fitted from experimental results. In this so-called relaxation time approximation (RTA), the Boltzmann equation in the linear limit is solved straightforwardly providing the traditional Drude model predictions with $\tau_{\rm eff} = \gamma^{-1}$. We here go beyond the RTA making use of the Landau weak coupling assumption \cite{Balescu}, where we neglect electron-electron and electron-phonon collisions with large deflection angle while retaining all grazing-angle contributions and recasting Eq. (\ref{BoltzEq}) into the so-called Fokker-Planck-Landau equation (FPLE) 

\begin{figure}[t]
\centering
\begin{center}
\includegraphics[width=0.5\textwidth]{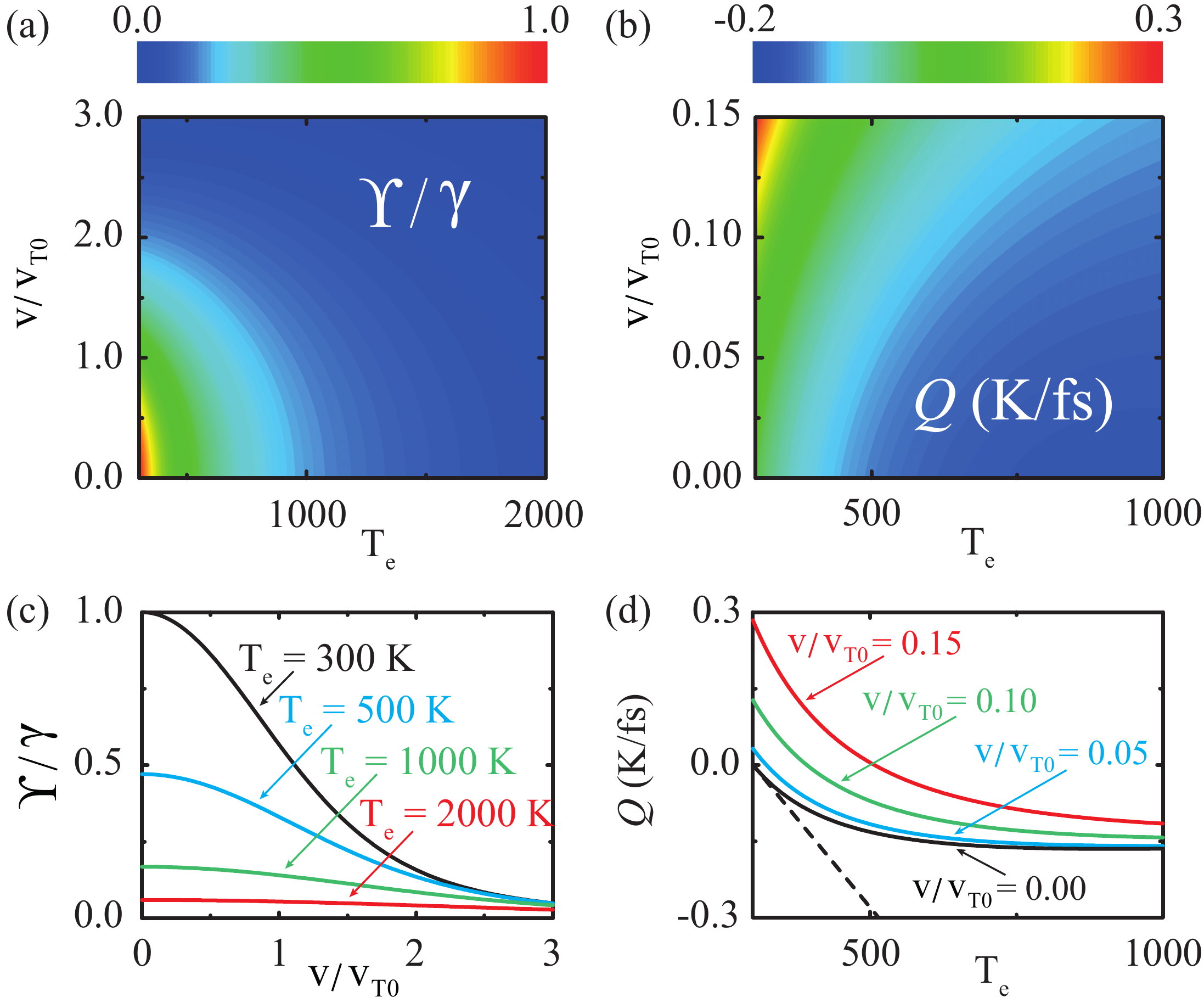}
\caption{{\bf Nonlinear heating and damping rates}. Parametric nonlinear dependency of {\bf (a,c)} rescaled nonlinear damping $\Upsilon/\gamma$ and {\bf (b,d)} heating rate $Q$ over the rescaled average velocity ${\rm v}/{\rm v}_{\rm T0}$ and OOE plasma temperature $T_{\rm e}$. The plots are made at ambient lattice temperature $T_0 = 300$ K and refer to silver, for which the constants in Eqs. (\ref{NlParEqs}) are $\omega_{\rm P} = 14$ fs$^{-1}$, $\gamma = 32$ ps$^{-1}$, and $\gamma_{\rm th} = 1.4$ ps$^{-1}$ \cite{CaiBook,Link1999}, obtained by setting $n_0 = 6.16 \times 10^{22}$ cm$^{-3}$, $m = 9.11 \times 10^{-28}$ g, $M = 4.07 \times 10^{-26}$ g, and ${\rm v}_{\rm T0} = 96.4$ km$/$s. The dashed black line in {\bf (d)} indicates the linear limit of plasma temperature relaxation $Q \simeq -\gamma_{\rm th} (T_{\rm e} - T_0)$ for vanishing velocity ${\rm v}\rightarrow 0$.}
\label{Fig1}
\end{center}
\end{figure}

\begin{subequations}
\label{FPLEqs}
\begin{eqnarray}
(\partial_tf)_{\rm coll}^{\rm el-el} & = & \frac{C_{\rm ee}}{m^2}\nabla_{\rm w}\cdot\{[ {\cal D}   \cdot\nabla_{\rm w} - 2             \nabla_{\rm w}\alpha   ]f\}, \label{FPLE1}\\
(\partial_tf)_{\rm coll}^{\rm el-ph} & = & \frac{C_{\rm ei}}{m^2}\nabla_{\rm w}\cdot\{[ {\cal D}_0 \cdot\nabla_{\rm w} - 2 \frac{m}{M} \nabla_{\rm w}\alpha_0 ]f\}, \label{FPLE2}
\end{eqnarray}
\end{subequations}
\noindent where $C_{\rm ee}$ and $C_{\rm ei}$ are two constants accounting for the electron-electron and electron-ion Coulomb interaction,  ${\cal D}(\beta) = \nabla_{\rm w}\nabla_{\rm w} \beta$ and ${\cal D}_{0} = {\cal D}(\beta_0)$ are diffusion tensors, and $\alpha({\bf r},{\bf w},t) = \int {\rm d}{\bf w}_1 f({\bf r},{\bf w}_1,t)/|{\bf w}-{\bf w}_1|$ and $\beta ({\bf r},{\bf w},t) = \int {\rm d}{\bf w}_1 f({\bf r},{\bf w}_1,t)|{\bf w}-{\bf w}_1|$ are the so-called Rosenbluth potentials \cite{Balescu,Rosenbluth} [Hereafter, $\alpha_0 = \alpha(f_0)$, $\beta_0 = \beta(f_0)$]. Borrowing such an approach, which is frequently used to precisely describe plasma dynamics in tokamaks \cite{Yoon2014}, in what follows we model electron dynamics in plasmonic materials beyond the RTA.

\paragraph{Hydrodynamical equations --} We solve Eq. (\ref{BoltzEq}) following the method of moments, which enables us to obtain a hierarchy of hydrodynamical equations that are fully equivalent to the FPLE \cite{Balescu}. Since we are interested only on the zero [electron density $\int {\rm d}{\bf w} f({\bf r},{\bf w},t) = n({\bf r},t)$, such that $\int {\rm d}{\bf r} n({\bf r},t) = N$], first [current density $\int {\rm d}{\bf w} {\bf w} f({\bf r},{\bf w},t) = n({\bf r},t){\bf v}({\bf r},t)$], and second [energy density $(m/2)\int {\rm d}{\bf w} |{\bf w} - {\bf v}|^2 f({\bf r},{\bf w},t) = (3/2) n({\bf r},t) k_{\rm B} T_{\rm e}({\bf r},t)$] moments, we truncate the hierarchy by neglecting higher-order moments obtaining the solution 
\begin{equation}
f({\bf r},{\bf w},t) \simeq \frac{n({\bf r},t) m^{3/2}}{[2\pi k_{\rm B}T_{\rm e}({\bf r},t)]^{3/2}} {\rm e}^{- \frac{\displaystyle m |{\bf w}-{\bf v}({\bf r},t)|^2}{\displaystyle 2k_{\rm B}T_{\rm e}({\bf r},t)}}. \label{TDFDDEQ}
\end{equation}

\begin{figure}[t]
\centering
\begin{center}
\includegraphics[width=0.5\textwidth]{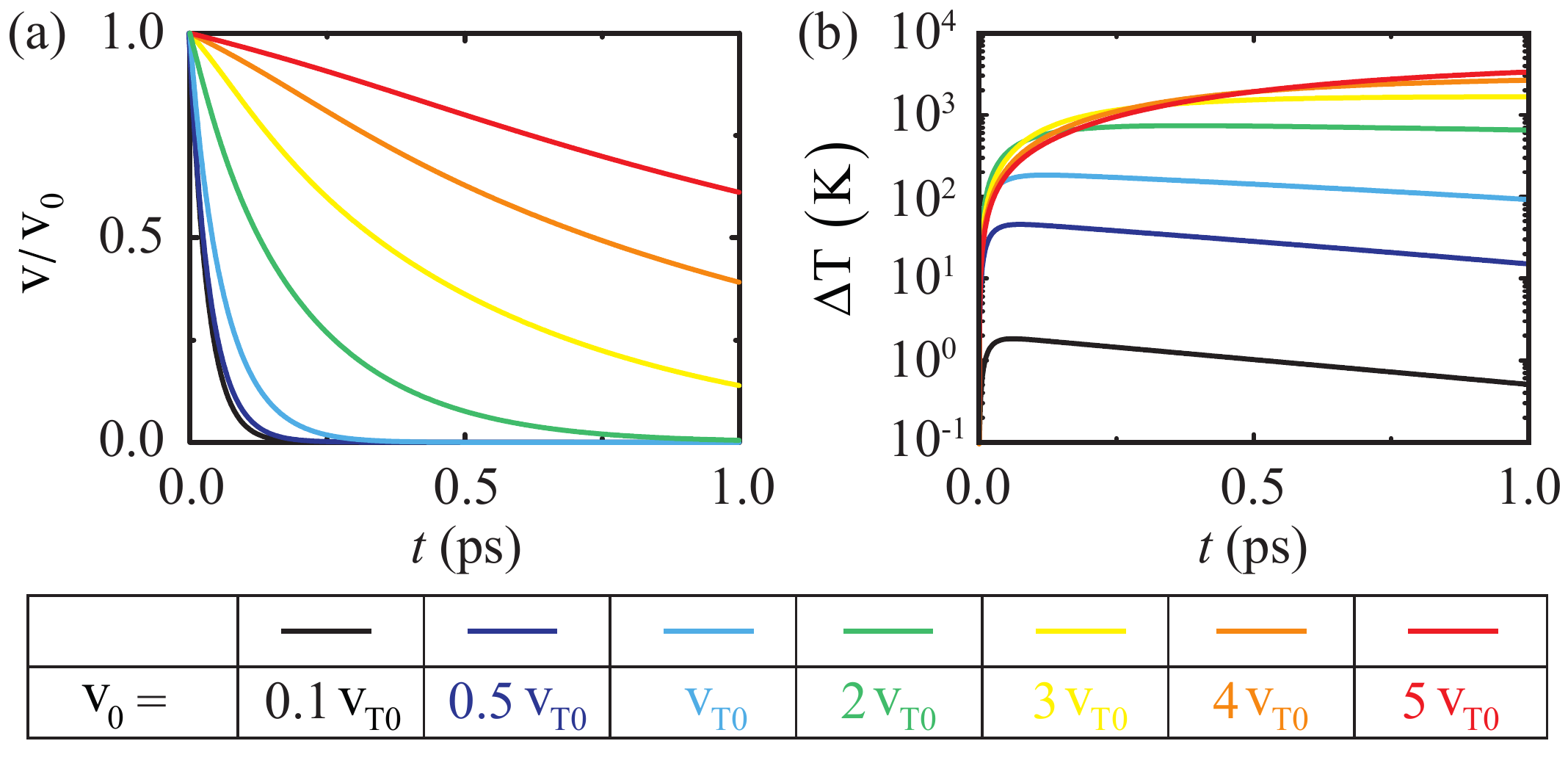}
\caption{{\bf Ballistic local response}. Evolution of {\bf (a)} rescaled average velocity ${\rm v}/{\rm v}_0$ and {\bf (b)} OOE temperature variation $\Delta T = T_{\rm e} - T_0$ for a plasma with initial temperature $T_0$ and velocities ${\rm v}_0$ indicated in the legend above. The plots illustrate calculations made for silver, with the same parameters used in Fig. 1.} 
\label{Fig2}
\end{center}
\end{figure}

\noindent The moments $n({\bf r},t)$, ${\bf v}({\bf r},t)$, and $T_{\rm e}({\bf r},t)$ satisfy the hydrodynamical equations (HDEs)
\begin{subequations}
\label{HDEQs}
\begin{eqnarray}
& & \partial_t n  + \nabla\cdot (n{\bf v}) = 0, \label{HDEQ1} \\
& & \partial_t {\bf v} + ({\bf v} \cdot \nabla){\bf v} + \frac{3k_{\rm B}}{mn}\nabla(n T_{\rm e}) = \frac{1}{m} {\bf F}_{\rm eff} - \Upsilon {\bf v}, \label{HDEQ2} \\
& & \partial_t T_{\rm e} + \frac{2}{3}T_{\rm e}\nabla\cdot{\bf v}+ {\bf v} \cdot \nabla T_{\rm e} = Q, \label{HDEQ3}
\end{eqnarray}
\end{subequations}
where
\begin{subequations}
\label{NlParEqs} 
\begin{eqnarray}
\Upsilon({\bf r},t) & = & \gamma \frac{3{\rm v}_{\rm T0}^3}{2{\rm v}^{2}{\rm v}_{\rm T}} \left[{\cal F}({\rm v}/{\rm v}_{\rm T}) - {\rm e}^{-\frac{\displaystyle{\rm v}^2}{\displaystyle{\rm v}_{\rm T}^2}}\right], \label{NLDampEq} \\
Q({\bf r},t)       & = & \gamma_{\rm th} \frac{M {\rm v}_{\rm T0}^3}{2 k_{\rm B}{\rm v}_{\rm T}}\left[ {\cal F}({\rm v}/{\rm v}_{\rm T}) - \frac{ T_{\rm e}{\rm v}_{\rm T0}^2}{T_0{\rm v}_{\rm T}^2}{\rm e}^{-\frac{\displaystyle{\rm v}^2}{\displaystyle{\rm v}_{\rm T}^2}}\right], \label{NLHeatEq}
\end{eqnarray}
\end{subequations}
${\bf F}_{\rm eff} = -e {\bf E} - e {\bf v}\times{\bf B}$ is the external effective force, ${\rm v}_{\rm T}(T_{\rm e}) = \sqrt{2k_{\rm B}(T_0/M+T_{\rm e}/m)}$ and ${\rm v}_{\rm T0} = {\rm v}_{\rm T}(T_0)$ are the OOE and equilibrium thermal velocities, respectively, $\gamma_{\rm th} = 2m\gamma/(m+M)$, $\gamma = 8\pi C_{\rm ei}n_0\sqrt{M/m(m+M)}/3(2\pi k_{\rm B}T_0)^{3/2}$, and ${\cal F}({\rm v}/{\rm v}_{\rm T}) = (\sqrt{\pi}{\rm v}_{\rm T}/2{\rm v}){\rm erf}(\rm v/{\rm v}_{\rm T})$. The current damping rate induced by electron collisions $\Upsilon$ and the heating rate $Q$ have been calculated by integrating analytically Eqs. (\ref{FPLE1},\ref{FPLE2}) making use of Eq. (\ref{TDFDDEQ}). Such quantities depend over time $t$ and position ${\bf r}$ only parametrically through the OOE macroscopic electron velocity ${\bf v}({\bf r},t)$ and temperature $T_{\rm e}({\bf r},t)$. The parametric functions $\Upsilon({\rm v},T_{\rm e})$ and $Q({\rm v},T_{\rm e})$ are depicted in Fig. \ref{Fig1} for silver. The novel HDEs reported above in Eqs. (\ref{HDEQs}) are the main result of this Letter and constitute a substantial extension of traditional HDEs used to model the EM response of plasmonic materials \cite{Scalora,CiraciPRB2012,KBusch}, where $\Upsilon \simeq \gamma$. Note that, in the limit of small electron current ${\rm v}/{\rm v}_{\rm T} <<1$ and electron heating $T_{\rm e}-T_0<<T_0$, Eq. (\ref{NLDampEq}) reduces to $\Upsilon \simeq \gamma$ (see Fig. \ref{Fig1}), thus recovering the RTA, while Eq. (\ref{NLHeatEq}) reduces to $Q \simeq - \gamma_{\rm th} (T_{\rm e}-T_0)$, which describes relaxation towards equilibrium via electron-phonon scattering. Conversely, our results indicate that damping $\Upsilon$ quenches for ${\rm v}/{\rm v}_{\rm T0}>>1$ (see Figs. \ref{Fig1}a,c). We emphasize that the damping rate $\Upsilon$ ensues only from $(\partial_t f)_{\rm coll}^{\rm el-ph}$ since electron-electron collisions conserve total momentum and energy \cite{Balescu}, implying that $(\partial_t f)_{\rm coll}^{\rm el-el} = 0$, which we obtain analytically. From a quantum perspective electron-electron collisions in metals can contribute to damping through an Umklapp process involving a reciprocal lattice vector, but such a process is quite inefficient in most EM frequency ranges and it becomes comparable with the electron-phonon scattering only at optical frequencies \cite{Khurgin2015}. 


\begin{figure}[t]
\centering
\begin{center}
\includegraphics[width=0.5\textwidth]{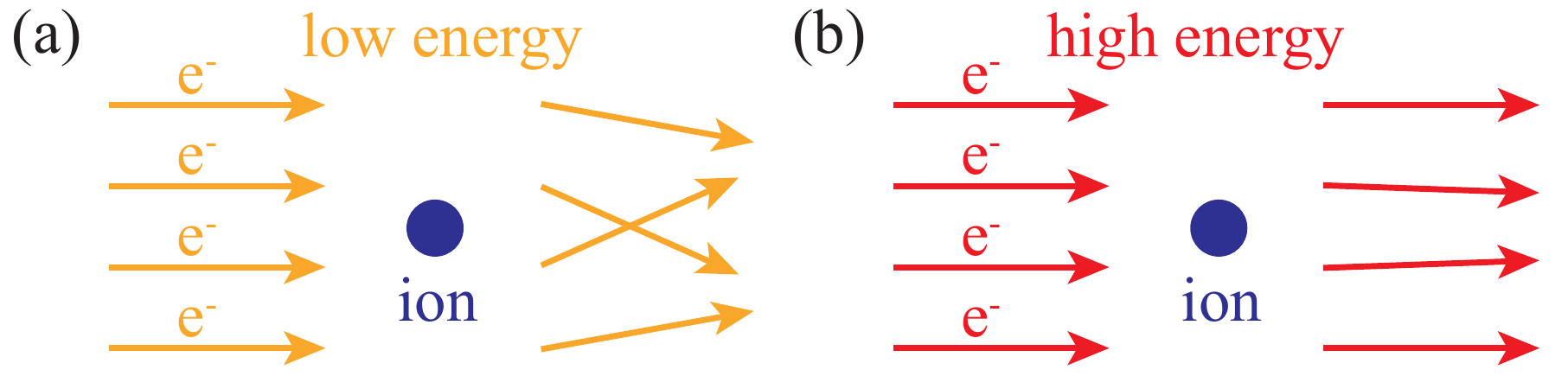}
\caption{{\bf Physical interpretation of damping quenching}. The trajectory deflection of {\bf (a)} low-energy electrons is much larger than for {\bf (b)} high-energy electrons owing to the reduced Coulomb interaction with ions. This effect mitigates electron damping and the related EM absorption.}
\label{Fig3}
\end{center}
\end{figure}

\begin{figure}[t]
\centering
\begin{center}
\includegraphics[width=0.5\textwidth]{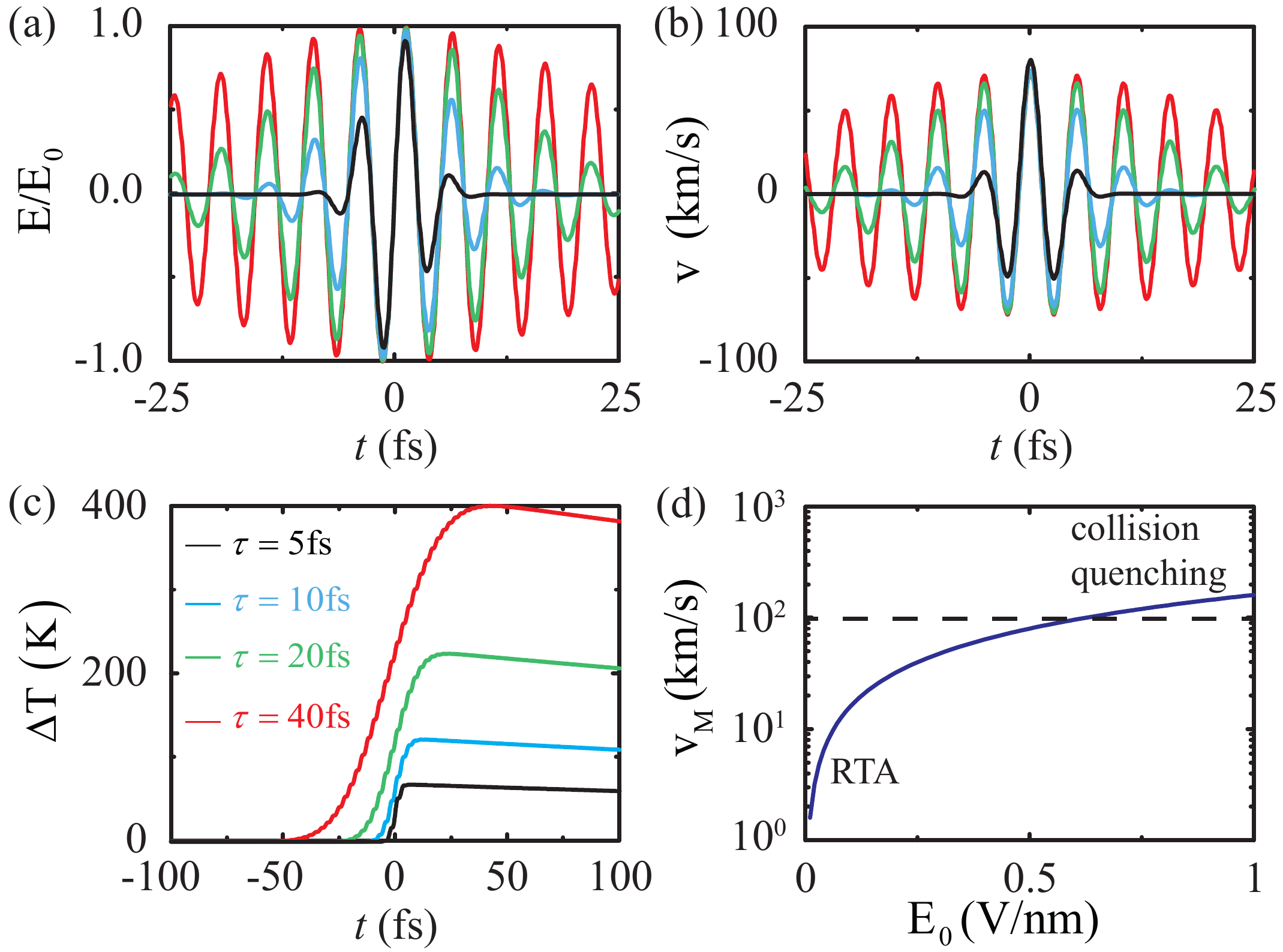}
\caption{{\bf Local response to ultrafast EM pulses}. Time evolution of plasma {\bf (b)} velocity ${\rm v}$ and {\bf (c)} temperature variation $\Delta T = T_{\rm e} - T_0$ for {\bf (a)} an ultrafast EM pulse with peak amplitude $E_0 = 5\times 10^8$ V$/$m (intensity $I_0 \simeq 3.32$ GW$/$cm$^2$), wavelength $\lambda = 1.55$ $\mu$m, and several FWHM $\tau$ indicated in the legend in {\bf (c)}. {\bf (d)} Dependence of the maximum velocity ${\rm v}_{\rm M}$ reached by electrons as a function of the amplitude of the driving EM pulse (no dependence over the pulse FWHM $\tau$ is observed). The black dashed line in {\bf (d)} indicates the equilibrium thermal velocity ${\rm v}_{\rm T0} = 96.4$ km$/$s at which damping quenching ignites. The plots illustrate calculations made for silver, as in Figs. 1,2.}
\label{Fig4}
\end{center}
\end{figure}

\paragraph{Local response --} The LHSs of Eqs. (\ref{HDEQs}), in addition to temporal evolution, account also for nonlocal contributions including Landau damping and temperature diffusion, which are well-known effects \cite{Scalora,CiraciPRB2012,KBusch}. In order to describe the novel ultrafast dynamics predicted by our model, we here focus on the local response, which first understanding can be grasped by initially ignoring the external driving field and focusing only on plasma relaxation. In Fig. \ref{Fig2} we plot the temporal evolution (obtained by numerical integration of Eqs. (\ref{HDEQs}) by a fourth-order Runge-Kutta algorithm) of (a) ${\rm v}(t)$ and (b) $\Delta T(t) = T_{\rm e}(t) - T_0$ for fixed initial temperature $T_{\rm e} (t=0) = T_0$ and several initial velocities ${\rm v} (t=0) = {\rm v}_0$. Note that when ${\rm v}_0$ overcomes ${\rm v}_{\rm T0}$ the damping rate towards equilibrium ${\rm v} = 0$ is heavily quenched (see Fig. \ref{Fig2}a). Intuitively, damping quenching for large macroscopic velocities can be understood as a reduced electron deflection upon collision (originating damping) owing to the increased kinetic energy, as pictorially illustrated in Fig. \ref{Fig3}. Note also that, as a consequence of damping quenching, $\Delta T(t)$ saturates at large ${\rm v}_0/{\rm v}_{\rm T0}$ due to the reduced rate of energy transfer (see Fig. \ref{Fig2}b). 

This intuitive understanding from ballistics holds also when considering ultrafast plasma dynamics upon excitation by an external driving field ${\bf E}(t) = E_0 {\rm e}^{-(2{\log2})t^2/\tau^2}{\rm sin}\omega t {\bf \hat x}$ with linear polarization along ${\bf \hat x}$, peak amplitude $E_0$, carrier angular frequency $\omega$ and intensity full width at half maximum (FWHM) $\tau$ (see Fig. \ref{Fig4}a). By solving Eqs. (\ref{HDEQs}) through a fourth-order Runge-Kutta algorithm we observe that for silver, at intensities $I_0 = (1/2)\epsilon_0 c |E_0|^2$ (where $c$ is the speed of light in vacuum) of the order of GW$/$cm$^2$ and infrared carrier wavelength $\lambda = 1.55$ $\mu$m, the plasma velocity can reach the thermal velocity ${\rm v}_{\rm T0} = 96.4$ km$/$s (see Figs. \ref{Fig4}b,d), thus implying that damping in the induced current ${\bf J}(t) = -n_0 e {\bf v}(t)$ is quenched and EM absorption accordingly. In addition, such a reduced absorption is accompanied by a very limited increase of plasma temperature $\Delta T (t)$ (see Fig. \ref{Fig4}c) for pulse durations of few tens of femtoseconds, which prevents from material damage. The frequency dependence of the saturation intensity $I_{\rm S}$ at which EM absorption quenches can be derived by assuming a monochromatic field ${\bf E}(t) = {\rm Re} [{\bf E}_0 {\rm e}^{-{\rm i}\omega t}]$ and solving Eqs. (\ref{HDEQs}) perturbatively by setting ${\bf v} \simeq {\rm Re} [ ({\bf v}_1 + {\bf v}_3 |E_0|^2) {\rm e}^{-{\rm i}\omega t}] + {\cal O}(|E_0|^4)$ and $T_{\rm e}\simeq T_0 + T_1 |E_0|^2+ {\cal O}(|E_0|^4)$, and by neglecting higher order harmonics and powers of $|E_0|^2$. We find that the plasmonic material polarization is ${\bf P} \simeq \epsilon_0 {\rm Re}[(\chi_{\rm Drude} + \chi_3 |E_0|^2){\bf E}_0{\rm e}^{-{\rm i}\omega t}]$, where $\chi_3 = - {\rm i} a \gamma\omega_{\rm P}^2(\omega-{\rm i}\gamma)^2/\omega(\omega^2+\gamma^2)^3$ and $a = Me^2(9m+10M)/40m^2(m+M)k_{\rm B}T_0$, while the absorption saturation intensity $I_{\rm sat} = - \epsilon_0 c {\rm Im}\chi_{\rm Drude}/4{\rm Im}\chi_3$  (${\rm Im}\chi_{\rm Drude}>0$, ${\rm Im}\chi_3<0$) depends quadratically over the angular frequency $\omega$ in the limit $\omega>>\gamma$. Thus, when the EM intensity is $I_0 = I_{\rm sat}$, the effective susceptibility $\chi_{\rm eff} = \chi_{\rm Drude} + \chi_3 |E_0|^2$ is such that ${\rm Im}\chi_{\rm eff} = {\rm Im}\chi_{\rm Drude}/2$ and the effective EM absorption coefficient is twice smaller than the linear one (for silver we obtain $I_{\rm S} \simeq 1.3$ GW$/$cm$^2$ at $\lambda = 1.55$ $\mu$m). Finally we emphasize that, besides noble metals in the infrared, by choosing different parameters our general model can be adapted to describe accurately ultrafast electron dynamics in any plasmonic material with purely intraband response, e.g., doped semiconductors with near-zero-index response such as indium tin oxide (ITO), aluminum zinc oxide (AZO), and indium zinc oxide (IZO), and similar results are expected for such materials as well as for laser-plasma interaction in air and in photonic crystal fibers filled with ionized gases.

\paragraph{Conclusions --}

We have developed a general theoretical framework to describe ultrafast dynamics of plasmas driven by pulses of EM radiation with duration of few femtoseconds and high peak intensity. Starting from the Boltzmann equation in the weak coupling assumption, we have developed a novel set of hydrodynamical equations able to describe the nonlinear dependencies of the plasma heating rate and damping, which quenches for large - but experimentally attainable - radiation peak intensities, leading to absorption saturation. Our results hold great potential for mitigating absorption of plasmonic materials, thus opening novel avenues for the development of low-loss plasmonic circuits \cite{Ozbay2006} and solid-state attosecond pulse sources \cite{Stockman} along with ultra-efficient nonlinear control at the nanoscale by near-zero index media \cite{Zahirul}. 

\


\noindent A.M. acknowledges support from the Rita Levi Montalcini Fellowship (grant number PGR15PCCQ5) funded by the Italian Ministry of Education, Universities and Research (MIUR), and from the Marie Curie Individual Fellowship OUTNANO (grant number 746774). A.C. acknowledges the U.S. Army International Technology Center Atlantic for financial support (grant number W911NF-14-1-0315). C.C. acknowledges the H2020 QuantERA Quomplex project (grant number 731743). A.M. acknowledges useful discussions with A. V. Zayats, C. Rizza and A. Puglisi.

\end{document}